\documentclass{elsart}
\usepackage{psfig}

\def\lsim{\mathrel{\rlap{\lower4pt\hbox{\hskip1pt$\sim$}}
   \raise1pt\hbox{$<$}}}         %less than or approx. symbol

\begin{document}
 
\begin{frontmatter}
 
\title{$a_0(980)$-$f_0(980)$ mixing and isospin violation in the
    reactions $pN \rightarrow d a_0$, $pd \rightarrow
    \mathrm{^3He/^3H}\, a_0$ and $ dd \rightarrow
    \mathrm{^4He}\, a_0$}

\author[INR]{V.Yu. Grishina\thanksref{DFG}},
\author[ITEP,Juelich]{L.A. Kondratyuk\thanksref{DFG}},
\author[Juelich]{M. B\"uscher\thanksref{DFG}\thanksref{author}},
\author[Giessen]{W. Cassing\thanksref{DFG}}, and
\author[Juelich]{H. Str\"oher}
\address[INR]{Institute for Nuclear Research, 60th October Anniversary
  Prospect 7A, 117312 Moscow, Russia}
\address[ITEP]{Institute of Theoretical and Experimental physics, B.\
  Cheremushkinskaya 25, 117259 Moscow, Russia}
\address[Juelich]{Institut f\"ur Kernphysik, Forschungszentrum
  J\"ulich, 52425 J\"ulich, Germany}
\address[Giessen]{Institute for Theoretical Physics, University of Giessen,
  D-35392 Giessen, Germany}
\thanks[DFG]{Supported by DFG and RFFI} 
\thanks[author]{Corresponding author; e-mail: m.buescher@fz-juelich.de} 
 
\begin{abstract}
  It is demonstrated that $f_0$-$a_0$ mixing can lead to a
  comparatively large isospin violation in the reactions $pN
  \rightarrow da_0$, $pd \rightarrow \mathrm{^3He/^3H}\, a_0$ and $dd
  \rightarrow \mathrm{^4He}\, a_0$ close to the corresponding production
  thresholds. The observation of such mixing effects is possible,
  e.g., by measuring the forward-backward
  asymmetry in the reaction $pn \rightarrow d a_0^0 \rightarrow
  d \eta \pi^0$.

\end{abstract}

\end{frontmatter}
 
As it was suggested long ago in Ref.~\cite{Achasov} the dynamical
interaction of the $a_0(980)$- and $f_0(980)$-mesons with states close
to the $K \bar K$ threshold may give rise to a significant
$a_0(980)$-$f_0(980)$ mixing. Different aspects of this mixing and the
underlying dymanics as well as the possibilities to measure this
effect have been discussed in
Refs.~\cite{Barnes,Achasov2,Janssen,Speth,Kerbikov}. Furthermore, it
has been suggested recently by Close and Kirk~\cite{Close} that the
new data from the WA102 collaboration at CERN~\cite{Barberis} on the
central production of $f_0$ and $a_0$ in the reaction $pp\rightarrow
p_s X p_f$ provide evidence for a significant $f_0$-$a_0$ mixing
intensity as large as $|\xi|^2=8\pm 3$\%.

In this letter we  discuss possible experimental tests of this 
mixing in the reactions 
$$pp\rightarrow da_0^+~~(a),~~pn \rightarrow da_0^0~~(b),$$
$$pd\rightarrow \mathrm{^3H}\, a_0^+~~(c),~~pd \rightarrow
\mathrm{^3He}\, a_0^0~~(d)$$ 
and
$$dd\rightarrow \mathrm{^4He}\, a_0^0~~(e)$$
near the corresponding
thresholds. We recall that the $a_0$-meson can decay to $\pi \eta$ or
$K \bar K$. In this paper we consider only the dominant $\pi \eta$
decay mode.

Note that the isospin violating anisotropy in the reaction $pn
\rightarrow da_0^0$ due to the $a_0(980)$-$f_0(980)$ mixing is very
similar to that what may arise from $\pi^0$-$\eta$ mixing in the
reaction $pn \rightarrow d \pi^0$ (see Ref.~\cite{Tippens}).  Recently
charge-symmetry breaking was investigated in the reactions $\pi^+ d
\rightarrow pp \eta$ and $\pi^- d \rightarrow nn \eta$ near the $\eta$
production threshold at BNL \cite{Tippens}.  A similar experiment,
comparing the reactions $p d \rightarrow \mathrm{^3 He} \pi^0$ and $p
d \rightarrow \mathrm{^3 H} \pi^+$ near the $\eta$ production
threshold, is now in progress at COSY-J\"ulich (see e.g.\ 
Ref.~\cite{Magiera}).

\section{Reactions (a) and (b)}

\subsection{Phenomenology of isospin violation}
\label{sec:pheno} 
In reactions (a) and (b) the final $da_0$ system has isospin $I_f=1$,
for $l_f=0$ ($S$-wave production close to threshold) it has spin-parity
$J^P_f=1^+$. The initial $NN$ system cannot be in
the state $I_i=1,~J^P_i=1^+$ due to the Pauli principle. Therefore,
near threshold the $da_0$ system should be dominantly produced in
$P$-wave with quantum numbers $J_f^P=0^-,~1^-$ or $2^-$. The states with
$J_i^P=0^-$, $1^-$ or $2^-$ can be formed by an $NN$ system with
spin $S_i=1$ and $l_i=1$ and 3. Neglecting the contribution of the
higher partial wave $l_i=3$ we can write the amplitude of reaction
(a) in the following form
\begin{eqnarray}
&&T(pp \rightarrow d~a_0^+)= \nonumber \\
&&=\alpha^{+}\
{\bf {p \cdot S}}\
{\bf {k}} \cdot \mbox{\boldmath $\epsilon$}{}^*
+\beta^{+}\
{\bf {p \cdot k}}\ {\bf {S}} \cdot \mbox{\boldmath $\epsilon$}{}^*
+\gamma^{+}\
{\bf {S \cdot k}}\ {\bf {p}} \cdot \mbox{\boldmath $\epsilon$}{}^*,
\end{eqnarray}
where ${\bf S}=\phi_N^T \sigma_2 \ \mbox{\boldmath$\sigma$}\phi_N$ is
the spin operator of the initial $NN$ system; $\bf{p}$ and $\bf{k}$
are the initial and final c.m.\ momenta; $\mbox{\boldmath $\epsilon$}$
is the deuteron polarization vector; $\alpha^+$, $\beta^+$, $\gamma^+$
are three independent scalar amplitudes which can be considered as
constants near threshold (for $k \rightarrow 0$).
 
Due to the mixing the $a_0^0$ may also be produced via the $f_0$. In
this case the $da_0^0$ system will be in $S$-wave and the
amplitude of reaction (b) can be written as:
\begin{eqnarray}
  &&T(pn \rightarrow d~a_0^0)= \nonumber \\
  &&=\alpha^{0}\ {\bf {p \cdot S}}\ {\bf {k}} \cdot \mbox{\boldmath
    $\epsilon$}{}^* +\beta^{0}\ {\bf {p \cdot k}}\ {\bf {S}} \cdot
  \mbox{\boldmath $\epsilon$}{}^* +\gamma^{0}\ {\bf {S \cdot k}}\ {\bf
    {p}} \cdot \mbox{\boldmath $\epsilon$}{}^* + \xi F \ \bf {S} \cdot
  \mbox{\boldmath $\epsilon$}{}^*,
\end{eqnarray}
where $\xi$ is the mixing parameter and $F$ is the $f_0$-production
amplitude. In the limit $k \rightarrow 0$, $F$ is again a constant.
The scalar amplitudes $\alpha$, $\beta$, $\gamma$ for reactions (a)
and (b) are related to each other by a factor $\sqrt{2}$, i.e.,
$\alpha^{+}=\sqrt{2} \alpha^0$, $\beta^{+}=\sqrt{2} \beta^0$,
$\gamma^{+}=\sqrt{2} \gamma^0$.
 
The differential cross sections for the reactions (a) and (b) have the
form (up to terms linear in $\xi$)
\begin{eqnarray}
  &\displaystyle \frac{{\mathrm{d}}\sigma(pp\rightarrow
    d~a_0^+)}{{\mathrm{d}}\Omega}=&
  2\ \frac{k}{p}\left(C_0+C_2 \cos^2 \Theta \right) \label{pp}\\
  &\displaystyle \frac{{\mathrm{d}}\sigma(pn\rightarrow d ~a_0^0)}
  {{\mathrm{d}}\Omega}=& \frac{k}{p}\left(C_0+ C_2 \cos^2 \Theta
  \right.+\left. C_1 \cos \Theta \right)\ ,\label{pn}
\end{eqnarray}
where
\begin{eqnarray}
\label{coeff} &&C_0=\frac{1}{2}\ p^2 k^2
\left[|\alpha^0|^2+|\gamma^0|^2 \right] ,~ C_1=p\ k
\left[\mathrm{Re} ((\xi F)^{*}(\alpha^0 +3\,
\beta^0+\gamma^0))\right] \ , \nonumber \\ &&C_2=\frac{1}{2}\ p^2
k^2\left[ 3\, |\beta^0|^2\right. \left. +2\, \mathrm{Re} (\alpha^0
\beta^{0\, *}+\alpha^0 \gamma^{0\, *}+ \beta^0 \gamma^{0\,
*})\right] \ .
\end{eqnarray}
Similarly, the differential cross section of the reaction $pn \rightarrow d
f_0$ can be written as
\begin{equation}
  \frac{{\mathrm{d}}\sigma( pn \rightarrow d f_0)}
  {{\mathrm{d}}\Omega}= \frac{3\, k}{2\, p}\ |F|^2\ . \nonumber
\end{equation}

The mixing effect --- described by the term $C_1\cos \Theta$ in
Eq.(\ref{pn}) --- then leads to an isospin violation in the ratio
$R_{ba}$ of the differential cross sections for reactions (b)
and (a),
\begin{equation}
R_{ba}=\frac12+\frac{C_1 \cos\Theta}{C_0+C_2\cos ^2\Theta} ,
\nonumber
\end{equation}
and in the forward-backward asymmetry for reaction (b):
\begin{equation}
A_a(\Theta)=\frac{\sigma_a(\Theta)-\sigma_a(\pi- \Theta)}
{\sigma_a(\Theta)+\sigma_a(\pi-\Theta)}=
\frac{C_1 \cos\Theta}{C_0+C_2\cos^2\Theta}\nonumber \ .
\label{asym}
\end{equation}

The latter effect was already discussed in Ref.~\cite{Kudryavtsev}
where it was argued that the asymmetry $A_a(\Theta=0)$ can reach
$5\div 10$\% at an energy excess of $Q=(5\div 10)$~MeV. However, if we
adopt a mixing parameter $|\xi|^2=(8\pm 3)$\%, as indicated by the
WA102 data, we can expect a much larger asymmetry. We note explicitly,
that the coefficient $C_1$ in (\ref{coeff}) depends not only on the
magnitude of the mixing parameter $\xi$, but also on the relative
phases with respect to the amplitudes of $f_0$ and $a_0$ production
which are unknown so far. This uncertainty has to be kept in mind for
the following discussion.

In case of very narrow $a_0$ and $f_0$ states, the differential cross
section (\ref{pp}), dominated by $P$-wave near threshold, would be
proportional to $k^3$ or $Q^{3/2}$, where $Q$ is the c.m.\ energy
excess. Due to $S$-wave dominance in the reaction $pn\rightarrow d
f_0$ one would expect that the cross section increases as $\sigma \sim
k $ or $\sim \sqrt{Q}$.  In this limit the $a_0$-$f_0$ mixing leads to
an enhancement of the asymmetry $A_a(\Theta)$ as $\sim 1/k$ near
threshold.  In reality, however, both $a_0$ and $f_0$ have a finite
width of about 40--100 MeV.  Therefore, at fixed initial momentum
their production cross section should be averaged over the
corresponding mass distributions, which will significantly change the
threshold behavior of the cross sections.  Another complication is
that broad resonances are usually accompanied by background lying
underneath the resonance signals.  These problems will be discussed
explicitly in Sects.~\ref{sec:model} and \ref{sec:bg}.

\subsection{Model calculations}
\label{sec:model} 
In order to estimate the isospin-violation effects in the ratio
$R_{ba}$ of the differential cross-section and in the forward-backward
asymmetry $A_a$ we use the two-step model (TSM), which has
successfully been applied to the description of $\eta$-,
$\eta^{\prime}$-, $\omega$- and $\phi$-meson production in the
reaction $pN \rightarrow d X$ in Refs.~\cite{Grishina1,Grishina2}.
Recently, this model has been also used for an analysis of the
reaction $pp \rightarrow d a_0^+$ \cite{Grishina3}.

\begin{figure}[t]
  \centerline{\psfig{file=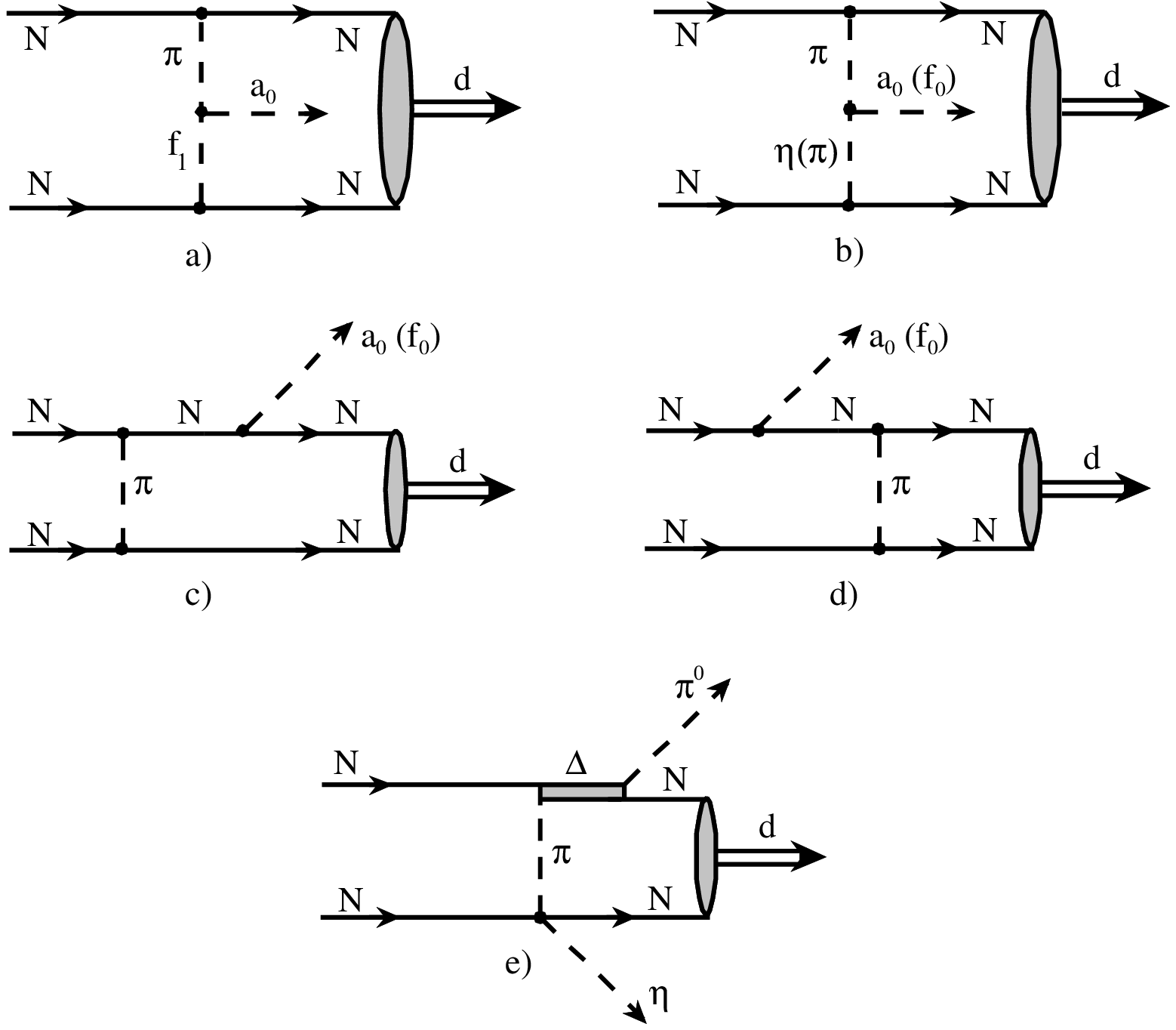,width=12cm}} 
  \caption{
    a)--d) Different mechanisms of $a_0$ and $f_0$-meson production in
    the reaction $NN\to da_0(f_0)$ within the framework of the
    two-step model (TSM). The nonresonant $\pi \eta$ production is
    described by diagram e).}
\label{fig:tsm}
\end{figure}

The diagrams in Fig.~\ref{fig:tsm} describe the different mechanisms
of $a_0$- and $f_0$-meson production in the reaction $NN\to da_0/f_0$
within the TSM. In the case of $a_0$ production the amplitude of the
subprocess $\pi N \to a_0 N$ contains three different contributions:
i) the $f_1(1285)$-meson exchange (Fig.~\ref{fig:tsm} a); ii) the
$\eta$-meson exchange (Fig.~\ref{fig:tsm}b); iii) $s$- and $u$-channel
nucleon exchanges (Fig.~\ref{fig:tsm}c and d). As it was shown in
Ref.~\cite{Grishina3} the main contribution to the cross section for
the reaction $pp \rightarrow d a_0^+$ stems from the $u$-channel
nucleon exchange (i.e.\ from the diagram of Fig.~\ref{fig:tsm}d and
all other contributions can be neglected in a leading order
approximation. In order to preserve the correct structure of the
amplitude under permutations of the initial nucleons (which is
antisymmetric for the isovector state and symmetric for the isoscalar
state) the amplitudes for $a_0$ and $f_0$ production can be written as
the following combinations of the $t$- and $u$-channel contributions
\begin{eqnarray}
&&T_{pn\to da_0^0}(s,t,u) = A_{pn\to da_0^0}(s,t)-A_{pn\to
  da_0^0}(s,u) \nonumber\\
&&T_{pn\to df_0}(s,t,u) = A_{pn\to df_0}(s,t)+A_{pn\to df_0}(s,u),
\label{Atu}
\end{eqnarray}
where $s=(p_1+p_2)^2$, $t=(p_3-p_1)^2$, $u=(p_3-p_2)^2$ and $p_1$,
$p_2$, $p_3$, and $p_4$ are the 4-momenta of the initial protons,
meson $M$ and the deuteron, respectively. The structure of the
amplitudes (\ref{Atu}) guarantees that the $S$-wave part vanishes in
the case of direct $a_0$ production since it is forbidden by
angular momentum conservation and the Pauli principle.
Also higher partial waves are included in the model calculations in
contrast to the simplified  discussion in Sect.~\ref{sec:pheno}.

In the case of $f_0$ production the amplitude of the subprocess $\pi N
\to f_0 N$ contains two different contributions: i) the $\pi$- meson
exchange (Fig.~\ref{fig:tsm} b); ii) $s$- and $u$-channel nucleon
exchanges (Fig.~\ref{fig:tsm} c and d). Our analysis has shown that
similarly to the case of $a_0$ production the main contribution to the
cross section of the reaction $pn \rightarrow d f_0$ is due to the
$u$-channel nucleon exchange (Fig.~\ref{fig:tsm} d); the contribution
of the combined $\pi\pi$ exchange (Fig.~\ref{fig:tsm} b) as well as
the $s$-channel nucleon exchange can be neglected.  In this case we
obtain for the ratio of the squared amplitudes
\begin{equation}
\frac{{|A_{pn\to df_0}}(s,t)|^2}{|A_{pn\to da_0}(s,t)|^2}=
\frac{{|A_{pn\to df_0}}(s,u)|^2}{|A_{pn\to da_0}(s,u)|^2}=
\frac{|g_{f_0NN}|^2}{|g_{a_0NN}|^2} .
\label{f0a0}
\end{equation}
If we take $g_{a_0NN}$ = 3.7 (see e.g.\ Ref.~\cite{Elster}) and
$g_{f_0NN}$ =8.5 \cite{Mull} then we find for the ratio of the
amplitudes $R(f_0/a_0) = g_{f_0NN}/g_{a_0NN} =2.3$. Note, however,
that Mull and Holinde give a different value for the ratio of the
coupling constants $R(f_0/a_0)=1.46$, which is about 37\% lower. In
the following we thus use $R(f_0/a_0)=1.46 \div 2.3$.

\begin{figure}[h]
\centerline{\psfig{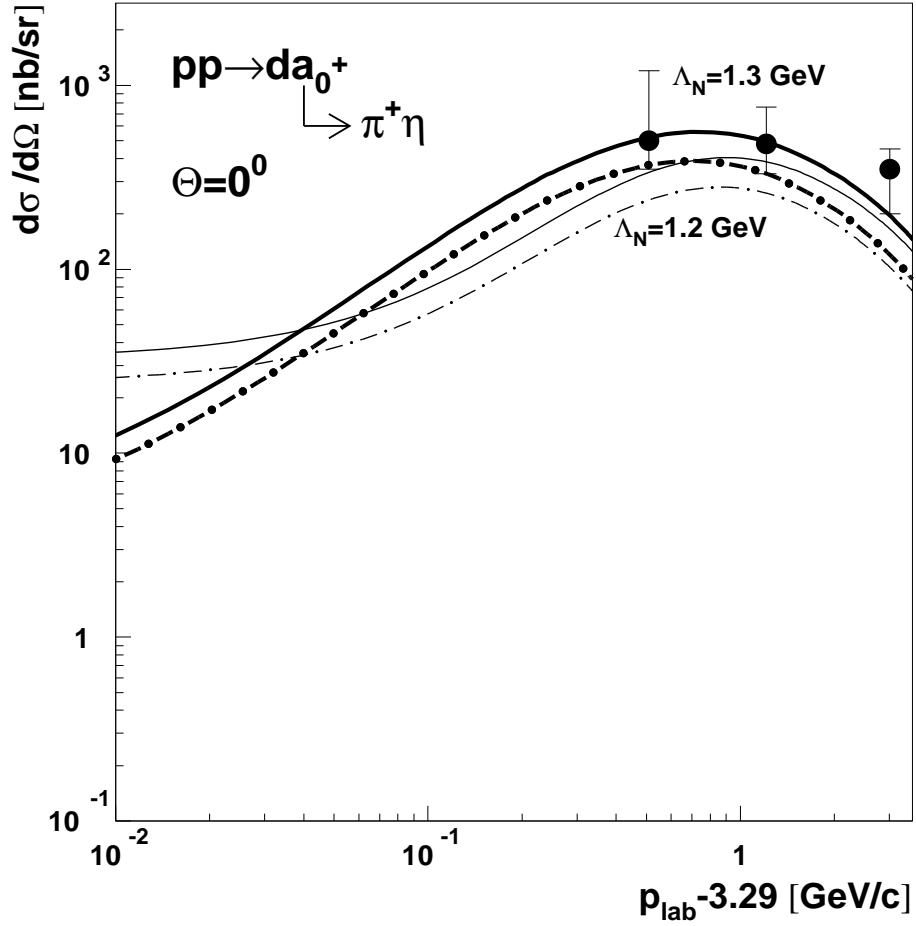}}
\caption{Forward differential cross section of the reaction
  $pp\rightarrow d a_0^+$ as a function of $(p_{\rm{lab}} - 3.29)$
  GeV/c. The full dots are the experimental data from
  Ref.~\protect\cite{Abolins} while the bold dash-dotted and solid
  lines describe the results of the TSM for $\Lambda_N = 1.2$ and 1.3
  GeV, respectively, and $\Gamma_{a_0}=0$.  The thin dash-dotted and
  solid lines are calculated using the Flatt\'e mass distribution for
  the $a_0$ spectral function with a cut $M \geq 0.85$ GeV (see
  text).}
\label{fig:dsacosy}
\end{figure}

The forward differential cross section for reaction (a) as a function
of the proton beam momentum is presented in Fig.~\ref{fig:dsacosy}.
The bold dash-dotted and solid lines (taken from Ref.~\cite{Grishina3}
and calculated for the zero width limit $\Gamma_{a_0}=0$) describe the
results of the TSM for different values of the nucleon cut-off
parameter, $\Lambda_N=1.2$ and 1.3 GeV/c, respectively.

In order to take into account the finite $a_0$ width we use a Flatt\'e
mass distribution with the same parameters as in Ref.~\cite{Brat2}:
K-matrix pole at 999 MeV, $\Gamma_{a_0 \to \pi \eta}=$ 70 MeV,
$\Gamma(K \bar K)/ \Gamma(\pi \eta) = 0.23$ (see also \cite{PDG} and
references therein).  The thin dash-dotted and solid lines in
Fig.~\ref{fig:dsacosy} are calculated within the TSM using this mass
distribution with a cut $M(\pi^+ \eta) \geq 0.85$ GeV and for
$\Lambda_N=1.2$ and 1.3 GeV, respectively.  The corresponding $\pi^0
\eta$ invariant mass distribution for the reaction $pn \rightarrow
da_0^0 \rightarrow d \pi^0 \eta$ at 3.4 GeV/c is shown in
Fig.~\ref{fig:MM} by the dashed line.

In case of the $f_0$, where the branching ratio $BR(K \bar K)$ is not
yet known \cite{PDG}, we use a Breit-Wigner mass distribution with
$m_R=980$ MeV and $\Gamma_R \simeq \Gamma_{f_0 \to \pi \pi}= 70$ MeV.

\begin{figure}[htb]
  \begin{center}
    \leavevmode
   \psfig{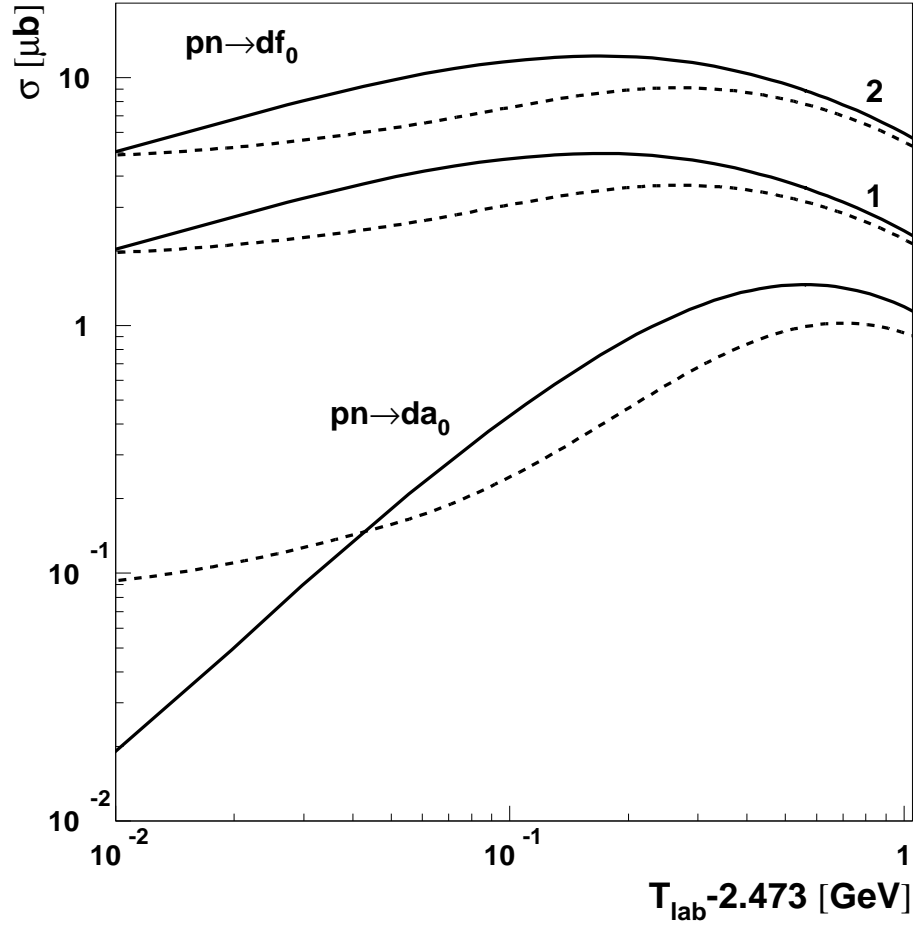}
    \caption{Total cross sections for the reactions
      $pn \to da_0$ (lower lines) and $pn \to df_0$ (upper lines) as a
      function of ($T_{\mathrm{lab}}-2.473$) GeV. The solid and dashed
      curves are calculated using narrow and finite resonance widths,
      respectively. The curves denoted by 1 and 2 correspond to the
      choices $R(f_0/a_0) = 1.46$ and 2.3. }
    \label{fig:siga0f0}
  \end{center}
\end{figure}

The calculated total cross sections for the reactions $pn \to da_0$
and $pn \to df_0$ (as a function of the beam energy $T_{\mathrm{lab}}$
for $\Lambda_N=1.2$ GeV ) are shown in Fig.~\ref{fig:siga0f0}. The
solid and dashed lines describe the calculations with zero and finite
widths, respectively. In case of $f_0$ production in the $\pi \pi$
decay mode we choose the same cut in the invariant mass of the $\pi
\pi $ system, i.e.\ $M_{\pi \pi} \geq 0.85$ GeV. The lines denoted by
1 and 2 are obtained for $R(f_0/a_0)=1.46$ and 2.3, respectively.
Comparing the solid and dashed lines it is obvious that near threshold
the finite width corrections to the cross sections are quite important
in particular for the energy behavior of the $a_0$-production cross
section (see also bold and thin curves in Fig.~\ref{fig:dsacosy}).

In principle, $a_0$-$f_0$ mixing can modify the mass spectrum of the
$a_0$ and $f_0$. However, in the $a_0$-$f_0$ case the effect is
expected to be less pronounced as for the $\rho$-$\omega$ case, where
the widths of $\rho$ and $\omega$ are very different (see e.g.\ the
discussion in Ref.~\cite{Tippens} and references therein).
Nevertheless, the modification of the $a_0^0$ spectral function due to
$a_0$-$f_0$ mixing can be measured by comparing the invariant mass
distributions of $a_0^0$ with that of $a_0^+$.  According to our
analysis, however, a much cleaner signal for isospin violation can be
obtained from the measurement of the forward-backward asymmetry in the
reaction $pn \to d a_0^0 \to d \pi^0 \eta$ integrating over the full
$a_0$ mass distribution. For the following calculations, the strengths
of the $a_0$ and $f_0$ thus will be integrated over the mass interval
0.85--1.02 GeV.

\begin{figure}[htb]
  \begin{center}
    \leavevmode
   \psfig{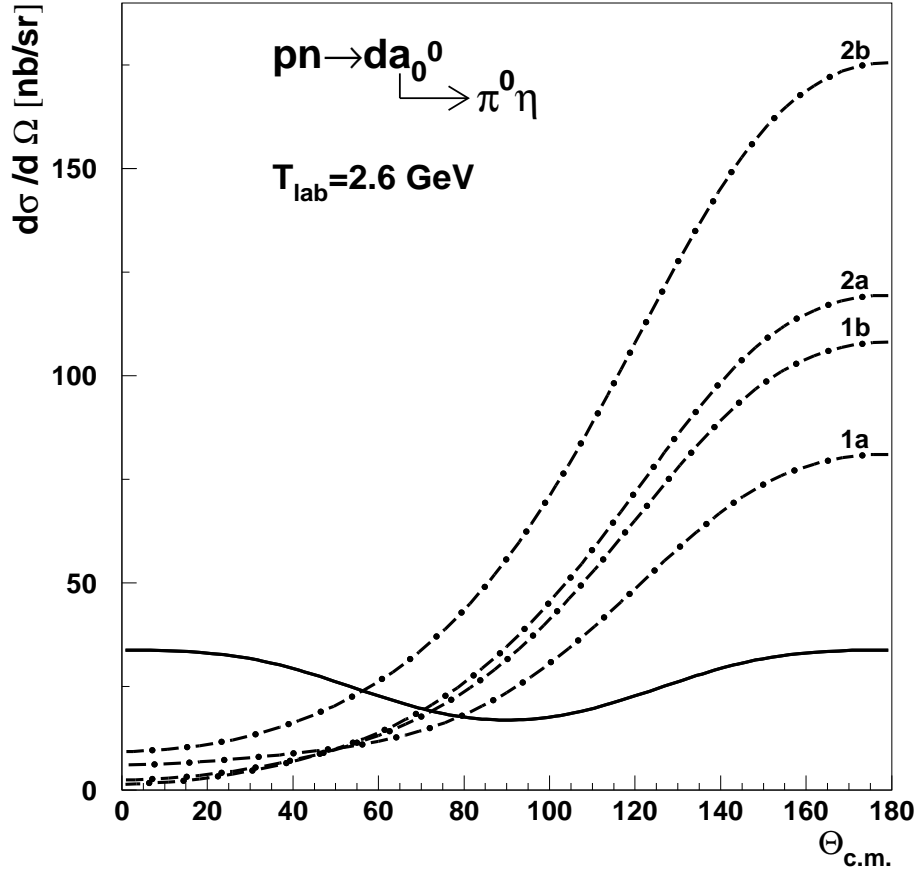}
    \caption{Differential cross section
      of the reaction $pn \to d a_0^0$ at $T_p=2.6$ GeV as a function
      of $\Theta_{\mathrm{c.m.}}$. The solid curve corresponds to the
      case of isospin conservation, i.e.\  $|\xi|^2=0$. The
      dashed-dotted lines include the mixing effect with $|\xi|^2 =
      0.05$ for the lower curves (1a and 2a) and $|\xi|^2 = 0.11$ for
      the upper curves (1b and 2b). The lines 1a, 1b (2a, 2b) have
      been calculated for $R(f_0/a_0) = 1.46$ (2.3), respectively.}
    \label{fig:dsa0f0}
  \end{center}
\end{figure}

The magnitude of the isospin violation effects is shown in
Fig.~\ref{fig:dsa0f0}, where we present the differential cross section
of the reaction $pn \to d a_0^0$ at $T_p=2.6$ GeV as a function of
$\Theta_{\mathrm{c.m.}}$ for different values of the mixing intensity
$|\xi|^2$ from 0.05 to 0.11. For reference, the solid line shows the
case of isospin conservation, i.e.\ $|\xi|^2=0$.  The dashed-dotted
curves include the mixing effect. Note that all curves in
Fig.~\ref{fig:dsa0f0} were calculated assuming maximal interference of
the amplitudes describing the direct $a_0$ production and its
production through the $f_0$.  The maximal values of the differential
cross section may also occur at $\Theta_{\mathrm{c.m.}}=0^{\circ}$
depending on the sign of the coefficient $C_1$ in Eq.(\ref{pn}).
 
It follows from Fig.~\ref{fig:dsa0f0} in either case that the
isospin-violation parameter $A_a(\Theta)$ for
$\Theta_{\mathrm{c.m.}}=180^{\circ}$ may be quite large, i.e.\ 
\begin{equation}
  A_a(180^\circ)= 0.86\div0.96~~ \mathrm{or}~~ 0.9\div0.98
\label{asymmax}
\end{equation}
for $R(f_0/a_0) = 1.46$ or 2.3, respectively.  Note that the asymmetry
depends rather weakly on $R(f_0/a_0)$.  It might be more sensitive to
the relative phase of $a_0$ and $f_0$ contributions, which has to be
settled experimentally.

\subsection{Background}
\label{sec:bg} 
\begin{figure}[htb]
  \begin{center}
    \leavevmode
   \psfig{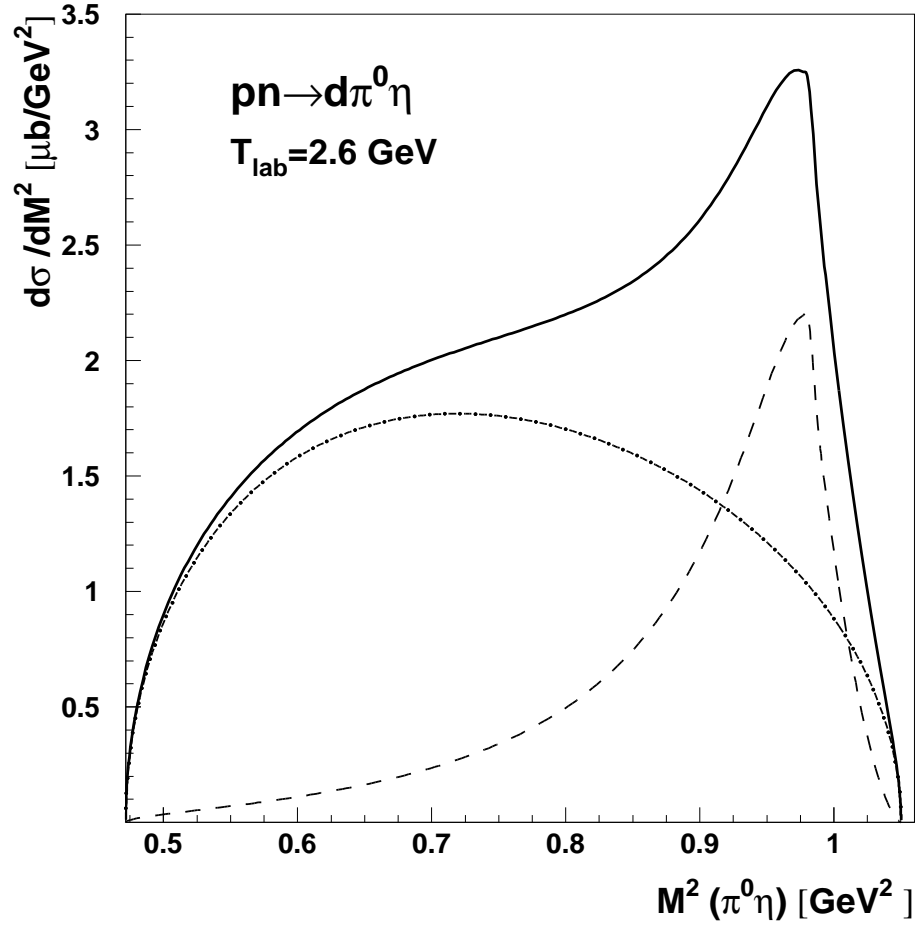}
    \caption{$\pi^0 \eta$ invariant mass distribution for the reaction
      $pn \rightarrow d \pi^0 \eta$ at 3.4 GeV/c. The dashed and
      dash-dotted lines describe the $a_0$-resonance contribution and
      nonresonant background, respectively. The solid line is the sum
      of both contributions.}
    \label{fig:MM}
  \end{center}
\end{figure}

The dash-dotted line in Fig.~\ref{fig:MM} shows our estimate of the
possible background from nonresonant $\pi^0 \eta$ production in the
reaction $pn \rightarrow d \pi^0 \eta$ at $T_{\mathrm{lab}}=2.6$ GeV
(see also Ref.~\cite{AnnRep2000}).  The background amplitude is
described by the diagram shown in Fig.~\ref{fig:tsm} e, where the
$\eta$ and $\pi$ mesons are created through the intermediate
production of a $\Delta(1232)$ (in the amplitude $\pi N \rightarrow
\pi N$) and a $N(1535)$ (in the amplitude $\pi N \rightarrow \eta N$).
The total cross section for the nonresonant $\pi \eta$ production due
to this mechanism was found to be $\sigma_{\mathrm{BG}} \simeq $ 0.8
$\mu$b for a cut-off in the one-pion exchange of $\Lambda_{\pi}= 1$
GeV.

We point out that the background is charge-symmetric and cancels in
the difference of the cross sections $\sigma(\Theta) - \sigma(\pi -
\Theta)$.  Therefore, a complete separation of the background is not
crucial for a test of isospin violation due to the $a_0$-$f_0$ mixing.
There will also be some contribution from $\pi$-$\eta$ mixing as
discussed in Refs.~\cite{Tippens,Magiera}. According to the results of
Ref.~\cite{Tippens} this mechanism yields a charge-symmetry breaking
in the $\eta NN$ system of about 6\%:
$$R=d\sigma (\pi^+d \rightarrow pp\eta)/\sigma (\pi^-d \rightarrow
nn\eta)=~0.938 \pm 0.009.$$
A similar isospin violation due to $\pi$-$\eta$ mixing can also be
expected in our case. 

The best strategy to search for isospin violation due to $a_0$-$f_0$
mixing is a measurement of the forward-backward asymmetry for
different intervals of $M_{\eta \pi^0}$.  It follows from
Fig.~\ref{fig:MM} that $\sigma_{a_0} (\sigma_{\mathrm{BG}})=
0.3(0.4),~0.27(0.29)$ and 0.19(0.15) $\mu$b for $M_{\eta \pi^0} \geq
0.85,~0.9$ and 0.95 GeV, respectively.  For $M_{\eta \pi^0} \leq 0.7$
GeV the resonant contribution is rather small and the charge-symmetry
breaking will dominantly be related to $\pi$-$\eta$ mixing and,
therefore, be small.  On the other hand, for $M \geq 0.95$ GeV the
background does not exceed the resonance contribution and we expect a
comparatively large isospin-breaking signal due to $a_0$-$f_0$ mixing.

\subsection{The reaction $pn \rightarrow d f_0 \rightarrow d \pi \pi $}
 
\begin{figure}[htb]
  \begin{center}
    \leavevmode
   \psfig{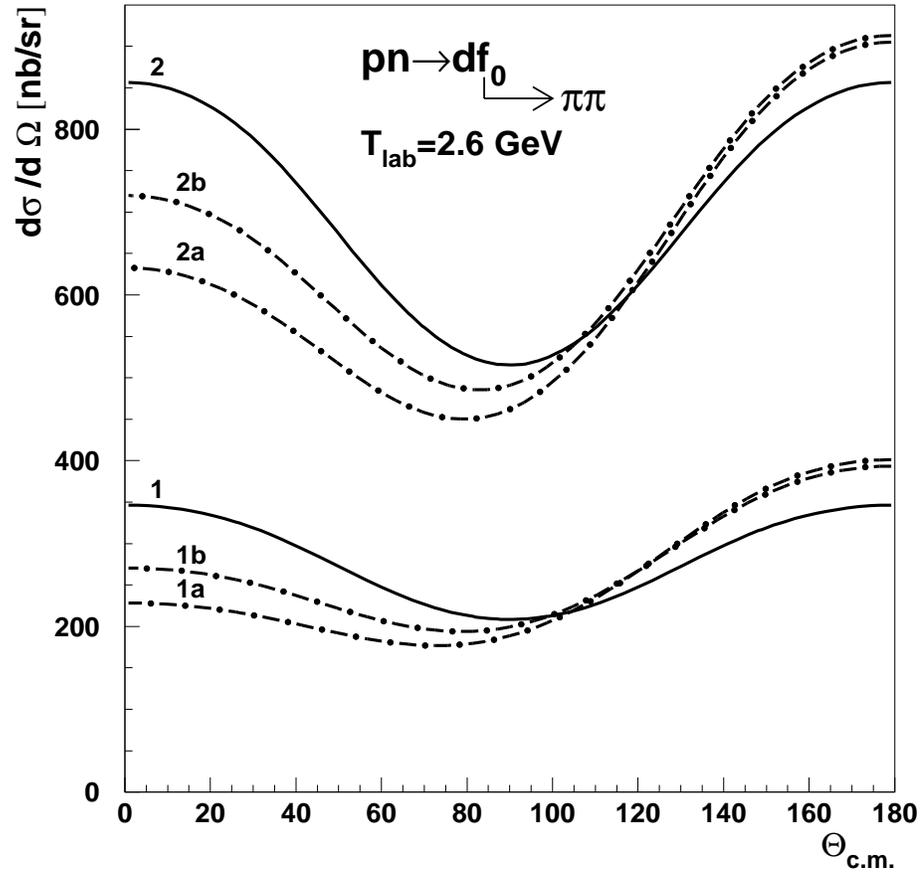}
    \caption{Differential cross section
      of the reaction $pn \to d f_0$ at $T_p=2.6$ GeV as a function of
      $\Theta_{\mathrm{c.m.}}$. The notation of the curves is the same
      as in Fig.~\ref{fig:dsa0f0}.}
    \label{fig:dsf0a0}
  \end{center}
\end{figure}

The isospin-violation effects can also be measured in the reaction
\begin{equation}
pn \to d f_0 \to d \pi^+ \pi^- , \label{f0pipi}
\end{equation}
where, due to mixing, the $f_0$ may also be produced via the $a_0$. 
The corresponding differential cross section is shown in
Fig.~\ref{fig:dsf0a0}. The differential cross section for $f_0$
production is expected to be substantiatially larger than for $a_0$
production, but the isospin violation effect turns out to be smaller
than in the $\pi \eta$-production channel. Nevertheless, the isospin
violation parameter $A$ is expected to be about 10$\div$30\% and can
be detected experimentally. 
 
\section{Reactions (c) and (d)}
We continue with $pd$ reactions and compare the final states
$\mathrm{^3H}\, a_0^+$ (c) and $\mathrm{^3He}\, a_0^0$ (d). Near
threshold the amplitudes of these reactions can be written as
\begin{eqnarray}
&&T(pd \rightarrow \mathrm{^3H}~ a_0^+)
=\sqrt{2} D_a \,{\bf {S_A}} \cdot \mbox{\boldmath $\epsilon$}
\end{eqnarray}
\begin{eqnarray}
&&T(pd \rightarrow \mathrm{^3He}~ a_0^0)=
(D_a+ \xi D_f)  \,{\bf {S_A}} \cdot \mbox{\boldmath $\epsilon$}\ ,
\end{eqnarray}
with ${\bf S_A}=\phi_A^T \sigma_2 \
\mbox{\boldmath$\sigma$}\phi_N$. Here $D_a$ and $D_f$ are the scalar
$S$-wave amplitudes describing the $a_0$ and $f_0$ production in case
of $\xi$=0. The ratio of the differential cross sections for
reactions (d) and (c) is then given by
\begin{equation}
R_{dc}=\frac{|D_a+ \xi D_f|^2}{2|D_a|^2} = \frac12+
\frac{2 \mathrm{Re}(D_a^* \xi D_f)+|\xi D_f|^2}{|D_a|^2}\ .
\label{R_dc}
\end{equation}
The magnitude of the ratio $R_{dc}$ now depends on the relative
value of the amplitudes $D_a$ and $D_f$. If they are comparable
$|D_a| \sim |D_f|$ or $|D_f|^2 \gg |D_a|^2$ the deviation of
$R_{dc}$ from 0.5 (which corresponds to isospin conservation)
might be 100\% or more. Only in the case $|D_f|^2 \ll |D_a|^2$ the
difference of $R_{dc}$ from 0.5 will be small. However, this
seems to be very unlikely.
 
Using the two-step model for the reactions $pd \rightarrow
\mathrm{^3He}~ a_0^0$ and $pd \rightarrow \mathrm{^3He}~ f_0$,
involving the subprocesses $pp\rightarrow d \pi^+$ and $\pi^+ n
\rightarrow p~ a_0/f_0$ (cf.\ Refs.~\cite{Faldt,Uzikov}), we find
\begin{equation}
\frac{\sigma(pd\rightarrow \mathrm{^3He}~a_0^0)}{\sigma(pd
\rightarrow\mathrm{^3He}~f_0)} \simeq
\frac{\sigma(\pi^+ n\rightarrow p~a_0^0)}{\sigma(\pi^+ n\rightarrow
  p~f_0)}\ .
\end{equation}
According to the calculations in Ref.~\cite{Grishina3} we expect
$\sigma(\pi^+ n \rightarrow pa_0)=\sigma(\pi^- p \rightarrow
na_0) \simeq 0.5\div 1$~mb at 1.75--2 GeV/c. A similar value for
$\sigma(\pi^- p \rightarrow nf_0)$ can be found using the results
from Ref.~\cite{Brat}. According to the latter study $\sigma(\pi^-
p\rightarrow nf_0 \rightarrow nK^+K^-) \simeq 6-8\ \mu$b at 1.75--2
GeV/c and $Br(f_0 \to K^+ K^-)\simeq 1\%$, which implies that
$\sigma(\pi^- p\rightarrow nf_0) \simeq 0.6-0.8$~mb. Thus we
expect that near threshold $|D_a| \sim |D_f|$ . This would imply
that the effect of isospin violation in the ratio $R_{dc}$ can
be rather large.

Recently the cross section of the reaction $pd\rightarrow
\mathrm{^3He}~K^+K^-$ has been measured by the MOMO collaboration at
COSY-J\"ulich. It was found that $\sigma =9.6 \pm 1.0$ and $17.5 \pm
1.8$ nb for $Q= 40$ and 56 MeV, respectively \cite{MOMO}.  The authors
note that the invariant $K^+K^-$ mass distributions in those data show
broad peaks which follow phase space.  However, as it was shown in
Ref.~\cite{Brat2}, the shape of an invariant mass spectrum following
phase space cannot be distinguished from an $a_0$-resonance
contribution at small values of $Q$.  Therefore, the events from
Ref.~\cite{MOMO} might also be attributed to $a_0$ and/or $f_0$
production.  Moreover, due to the phase space behavior near threshold
one expects a dominance of two-body reactions. Thus the cross section
of the reaction $pd \rightarrow \mathrm{^3He}~a_0^0 \rightarrow
\mathrm{^3He}~\pi^0 \eta$ is expected to be not significantly smaller
than the upper limit of about 80$\div$150~nb at $Q = 40-60$ MeV which
follows from the MOMO data (using $\Gamma(K \bar K)/ \Gamma(\pi \eta)
= 0.23$ from \cite{PDG}).

\section{Reaction (e)}
Any direct production of the $a_0$ in the reaction $dd \rightarrow
\mathrm{^4He}\, a_0^0$ is forbidden. It thus can only be observed due
to $f_0$-$a_0$ mixing:
\begin{equation}
\frac{\sigma(dd\rightarrow \mathrm{^4He}\, a_0^0)}
{\sigma(dd\rightarrow \mathrm{^4He}\, f_0)}= |\xi|^2.
\end{equation}
Therefore, it will be very interesting to study the reaction
\begin{equation}
dd \rightarrow \mathrm{^4He}\, (\pi^0~ \eta) \label{dd}
\end{equation}
near the $f_0$-production threshold. Any signal of the reaction
(\ref{dd}) then will be related to isospin breaking. It is
expected to be much more pronounced near the $f_0$ threshold as
compared to the region below this threshold.
 
\section{Summary}
In summary, we have discussed the effects of isospin violation in the
reactions $pN \rightarrow da_0$, $pd \rightarrow \mathrm{^3He/^3H}\,
a_0$ and $ dd \rightarrow \mathrm{^4He}\, a_0^0$ which can be generated
by $f_0$-$a_0$ mixing. It has been demonstrated that for a mixing
intensity of about ($8\pm3$)\%, the isospin violation in the ratio of
the differential cross sections of the reactions $pp \to da_0^+ \to d
\pi^+ \eta$ and $pn \to da_0^0 \to d \pi^0 \eta$ as well as in the
forward-backward asymmetry in the reaction $pn \to da_0^0 \to d \pi^0
\eta$ not far from threshold may be about 50--100\%.  Such large
effects originate from the interference of direct $a_0$ production and
its production via the $f_0$. The former amplitude is suppressed close
to threshold due to the $P$-wave amplitude whereas the latter is large
due to $S$-wave production.  A similar isospin violation is
expected in the ratio of the differential cross sections of the
reactions $pd \rightarrow \mathrm{^3H}\, a_0^+(\pi^+ \eta)$ and $pd
\rightarrow \mathrm{^3He}\, a_0^0(\pi^0\eta)$.  

Finally, we have also discussed the isospin-violation effects in the
reactions $pn \to df_0(\pi^+\pi^-)$ and $ dd \rightarrow
\mathrm{^4He}\, a_0$.  All reactions together --- once studied
experimentally --- are expected to provide detailed information on the
strength of the $f_0$-$a_0$ mixing.

Corresponding measurements are now in preparation for the ANKE
spectrometer at COSY-J\"ulich \cite{proposal}.

\newpage

\clearpage

\clearpage
 
\clearpage

\clearpage

\clearpage

\end{document}